\documentclass[intlimits,twoside,a4paper]{article}

\usepackage[cp1251]{inputenc}

\usepackage[eqsecnum]{cmpj3}

\usepackage{bm}

\issue{2024}{27}{2}{23702}
\doinumber{10.5488/CMP.27.23702}

\title[The local structure, electronic and optical properties of Pb(Mg$_{1/3}$Nb$_{2/3}$)O$_3$-PbTiO$_3$]%
{The local structure, electronic and optical properties of Pb(Mg$_{1/3}$Nb$_{2/3}$)O$_3$-PbTiO$_3$: first-principles study
}
 \author[M. Kovalenko, O. Bovgyra, V. Kapustianyk, O. Kozachenko]{M. Kovalenko\orcid{0000-0002-6848-6719}\thanks{Corresponding author:\email{mariya.kovalenko@lnu.edu.ua}.},  
   O. Bovgyra\orcid{0000-0001-9844-6868}, V. Kapustianyk\orcid{0000-0001-7830-5670}, O. Kozachenko\orcid{0000-0002-9747-6413}}
 \address{Solid State Physics Department, Faculty of Physics, Ivan Franko National University of Lviv, 8 Kyrylo and Mefodiy Str., 79005 Lviv, Ukraine}


\date{Received September 23, 2023, in final form November 02, 2023}

\begin{document}

\maketitle

\begin{abstract}
Pb(Mg$_{1/3}$Nb$_{2/3}$)O$_3$-PbTiO$_3$ perovskite-based crystals attract considerable scientific interest due to their inte\-res\-ting properties and possible use in piezoelectricity and photovoltaics. To understand the local structure and fundamental properties of such materials, in this work, we focused on the study within the density functional theory of structural, electronic, and optical properties of Pb[(Mg$_{1/3}$Nb$_{2/3}$)$_{0.75}$Ti$_{0.25}$]O$_3$. Using GGA(PBEsol) approximation for structure optimization gives a good agreement with experimental data. Through the variation in Hubbard $U$ parameters to GGA(PBEsol) functional, we achieve the bandgap for the Pb[(Mg$_{1/3}$Nb$_{2/3}$)$_{0.75}$Ti$_{0.25}$]O$_3$ which is in good agreement with the experimental results. The study of the bond populations showed that the Mg--O bond demonstrates no covalency, whereas there is a significant Ti--O and Nb--O covalent bonding. Such different bonding characteristics must be responsible for the relaxor properties of Pb[(Mg$_{1/3}$Nb$_{2/3}$)$_{0.75}$Ti$_{0.25}$]O$_3$ compound. In addition, the investigations of the optical properties of the Pb[(Mg$_{1/3}$Nb$_{2/3}$)$_{0.75}$Ti$_{0.25}$]O$_3$ by adopting Hubbard $U$ corrections, modifying the error of the GGA approximation, and confirming the electronic analysis, were performed.
%
%
\printkeywords
%
\end{abstract}

\section{Introduction}


Perovskite-based single crystals of lead magnesium niobate-lead titanate ($1-{x}$)Pb(Mg$_{1/3}$Nb$_{2/3}$)O$_3$--\textit{x}PbTiO$_3$ solid solution (PMN--PT) in which the Mg$^{2+}$ and Nb$^{5+}$ ions occupying the B-sites are initially disordered, are known as a relaxor ferroelectric. In essence, these dielectrics exhibit a wide, frequency-dependent response concerning temperature variations. These PMN--PT solid solutions create a new generation of piezoelectric materials because of their high piezoelectric coefficients ($d_{33} \sim 2500$~pC/N) and electromechanical coupling factors ($k_{33} \sim 94$\%) \cite{Luo2000}. Moreover, PMN--PT materials exhibit pronouncedly high dielectric constants and correspondingly low dielectric losses \cite{Kutnjak2006, Alguero2006}. These favorable characteristics significantly improve bandwidth and sensitivity when utilized in electromechanical sensing and power applications \cite{Bokov2000}. The outstanding piezoelectric properties of these materials manifest in compositions located within the morphotropic phase boundary (MPB) region, particularly close to the boundaries between the rhombohedral and monoclinic phases and the monoclinic and tetragonal phases. This is attributed to multiple polarization states or dipole orientations in materials with MPB compositions, which exhibit heightened susceptibility to electric field-driven switching. Consequently, these materials become more electrically active, significantly enhancing their piezoelectric response \cite{Noheda2002,  Ye2002}.

Another promising application area of the PMN--PT compound is its use in photovoltaic (PV) devices, such as semiconductor-based PV cells. In the study \cite{Semak2023}, the PMN--PT crystal's PV properties and correlation with its deformation properties were investigated. In addition, several ferroelectric compounds from the PMN--PT family became photovoltaic after WO$_3$ doping \cite{Tu2006} and the PV effect was reported for two compounds with stoichiometry near the MPB region \cite{Liew2022, Makh2019, Makh2018}.
Considering the above, these materials attract considerable scientific interest. This is confirmed by numerous experimental studies devoted to exploring dielectric dispersion and piezoelectric characteristics \cite{Bokov2002, Kutnjak2006, Shvar2013, Li2019, Li2021}. However, the precise relationship between composition and relaxor behavior still needs to be understood despite acknowledging the critical role played by hetero-valency or the degree of disorder on the B-site. Drawing from the existing research on relaxor ferroelectrics, there is a compelling proposition that the presence of local structural heterogeneity exerts a profound influence on the piezoelectric properties of these materials.

PMN--PT solid solutions are suitable objects for studying the correlation between structure and properties due to the availability of sufficient experimental data. However, to date there are only a few studies carried out within the density functional theory (DFT), which focused on the relationship between the local structure and electronic properties of such compounds \cite{Tan2018, Li2020, Grin2004, Takenaka2014}. Therefore, in this research we studied the influence of the local atomic environment on the electronic and optical properties of PMN--PT solid solutions using the first-principles methods within DFT.

\section{Model and methods}

To study the local structure of PMN--PT, first-principle calculations within DFT were carried out for a $2 \times 2 \times 2$ supercell model system of Pb[(Mg$_{1/3}$Nb$_{2/3}$)$_{0.75}$Ti$_{0.25}$]O$_3$ (0.75PMN--0.25PT) containing 40~atoms which was successfully used in previous studies \cite{Tan2018,Li2020}. This composition and B-cation ordering is a highly convenient alternative for first-principles investigations due to its capability of applying a relatively compact supercell. As a result, we adopt this specific system as the primary model for our research calculations. Furthermore, that supercell is set to be large enough to cover all changes in the local structure~\cite{Grin2004, Grin2005, Grinb2004}. Periodic calculations within ab initio DFT were performed using the pseudopotential plane wave method implemented in the CASTEP software code \cite{Clark2005}. The exchange-correlation functional is presented in the generalized gradient approximation (GGA) in the PBE parameterization for solids (PBEsol). The cut-off energy is set at 600 eV for all calculations. Representation of electronic states in the first Brillouin zone using a 2 $\times$ 2 $\times$ 2 $k$-points grid was performed according to the Monkhorst--Pack scheme. Equilibrium crystal structures were obtained by geometry optimization in the Broyden--Fletcher--Goldfarb--Shanno~(BFGS) minimization algorithm until the forces become less than 0.005~eV$\cdot$\AA$^{-1}$. To accurately describe the electronic structure of the 0.75PMN--0.25PT solid solution, the Hubbard $U$ correction method for the GGA(PBEsol) approximation was additionally used. This method is effectively used to describe the electronic structure of various systems of different dimensions \cite{Bovgyra2015, Bovgyra2016, Bovgyra2023, Kapus2022}.

\section{Results and discussion}

\begin{figure}[htb]
\centerline{\includegraphics[width=0.65\textwidth]{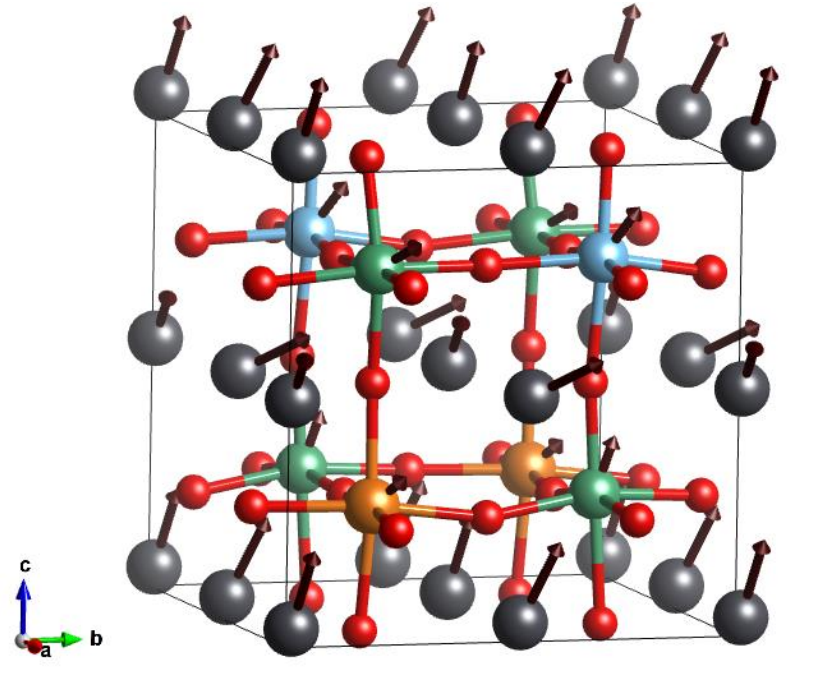}}
\caption{(Colour online) Optimized supercell structure of 0.75PMN--0.25PT received by DFT calculations. Arrows (with a scale factor of 3 for better visualization) indicate the displacements from high-symmetry perovskite positions. Gray, green, orange, blue, and red represent Pb, Nb, Mg, Ti, and O atoms.} \label{Fig1}
\end{figure}

Figure~\ref{Fig1} presents the 0.75PMN--0.25PT structure after geometry relaxation. The structure of PMN--PT corresponds to Pb(B$'_{1/2}$B$''_{1/2}$)O$_3$ \cite{Davies2000} for which the arrangement of B cations follows the model of random positions and for Nb atoms occupy the B$ ' $ cation site, and the B$ '' $ site Mg and Ti atoms fill equally. The optimized lattice parameters for the 0.75PMN--0.25PT supercell are $a = 8.095$~\AA, $b = 7.962$~\AA, $c = 8.177$~\AA, $\alpha = 90.14^\circ$, $\beta  = 89.34^\circ$, $\gamma  = 90.00^\circ$ and volume $V = 527.058$~\AA$^3$ corresponding to 65.88~\AA$^3$/f.u. The obtained value for volume per formula unit is in good agreement with the experimental data that gives approximately 64.88~\AA$^3$/f.u. \cite{Sepl2011}. 

The Pb ions and B-cations displacements, particularly Mg, Nb, and Ti ions inside the corresponding oxygen shells, determine macroscopic properties, especially for polarization in ferroelectric perovskites based on PMN--PT. 6\textit{s}-electrons of the lone pair of Pb\textsuperscript{2+} ions cause a significant displacement of Pb ions and the crucial
contribution of these ions to the overall polarization \cite{Makh2022}. The results of DFT calculations for the average displacement values of Pb\textsuperscript{2+}, Mg\textsuperscript{2+}, Nb\textsuperscript{5+}, and Ti\textsuperscript{4+} cations from the center of oxygen cage in the 0.75PMN--0.25PT solid solution are presented in table~\ref{tab1}, and are depicted by arrows in figure~\ref{Fig1}.

The local structure analysis in the equilibrium state demonstrates that the Pb ions undergo the most significant displacement of 0.446~\AA~from their initial positions, moving towards the Mg--Nb surface {[}(001) plane{]} and bypassing the Ti--Nb surface {[}(001) plane{]}, because the repulsive force between Pb--Mg atoms is weaker than that between Pb--Nb and Pb--Ti atoms. Nb and Ti ions move in the same direction as Pb, while Mg ions are only slightly shifted. The analysis shows that Nb and Ti atoms move from the center by 0.21--0.25~\AA, leading to shorter bonds between Nb--O and Ti--O while making a decisive contribution to macroscopic polarization \cite{Grinb2007}. The calculated local structure parameters agree well with previous theoretical and experimental results (see table~\ref{tab1}) \cite{Semak2023, Tan2018, Li2020, Sepl2011}.

\begin{table}[htb]
\caption{Local structure parameters of the 0.75PMN--0.25PT system: $V$ is
the volume of the $2 \times 2 \times 2$ supercell (in~\AA$^{3}$),
$D$\textsubscript{Pb}, $D$\textsubscript{Mg}, $D$\textsubscript{Nb}, and
$D$\textsubscript{Ti} are the average displacements of
Pb\textsuperscript{2+}, Mg\textsuperscript{2+}, Nb\textsuperscript{5+}, and Ti\textsuperscript{4+} cations, respectively (in~\AA).}
\label{tab1}
\vspace{2ex}
\begin{center}
\renewcommand{\arraystretch}{0}
\begin{tabular}{|c|c|c|c|c|c|}
\hline
Method & $V$ &$D$\textsubscript{Pb}&$D$\textsubscript{Mg}&$D$\textsubscript{Nb}&$D$\textsubscript{Ti}\strut\\
\hline
 GGA(PBEsol) (our data) & 527.058 & 0.446 & 0.078 & 0.207 & 0.252 \strut\\
 \hline
LDA \cite{Tan2018} & 509.1 & 0.368 & 0.066 & 0.173 & 0.233 \strut\\
\hline
GGA(PBE) \cite{Semak2023} & 510.431 & 0.395 & 0.076 & 0.177 & 0.220 \strut\\
\hline
LDA \cite{Grin2004} & -- & 0.389 & 0.080 & 0.181 &
0.181 \strut\\
\hline
\end{tabular}
\renewcommand{\arraystretch}{1}
\end{center}
\end{table}

\begin{figure}[htb]
\centerline{\includegraphics[width=0.75\textwidth]{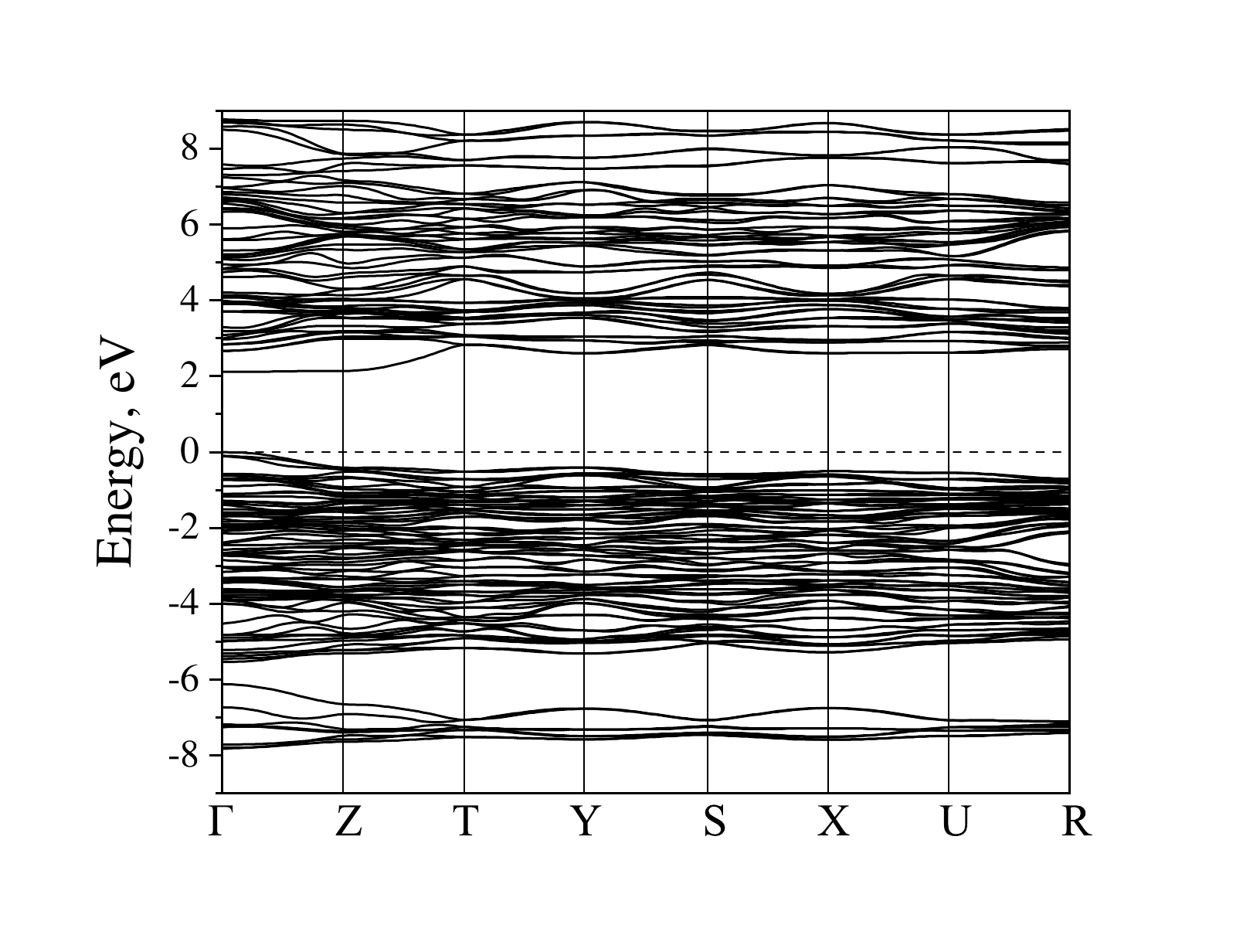}}
\caption{Band structure of the 0.75PMN--0.25PT, calculated by the GGA(PBEsol) method. The horizontal dashed line indicates the Fermi level.} \label{fig2}
\end{figure}

After structure optimization, the electronic structure of 0.75PMN--0.25PT solid solution was studied. Figure~\ref{fig2} presents the band structure of 0.75PMN--0.25PT, calculated along the high symmetry points of the first Brillouin zone. The bandgap (${E}_{g}$) calculated using the GGA(PBEsol) at the $\Gamma$ point is 2.13~eV. This value is larger than for the PbTiO\textsubscript{3} crystal (1.88~eV \cite{Zhang2017}), calculated by the same method.  The obtained value of ${E}_{g}$ is smaller compared to the experimental one (3.24~eV \cite{Wan2004}), and such underestimation is a common problem of the GGA exchange-correlation functional. The results of solving this problem using the on-site Hubbard corrections (DFT+U method) will be presented herein below.

\begin{figure}[h!]
\centerline{\includegraphics[width=0.65\textwidth]{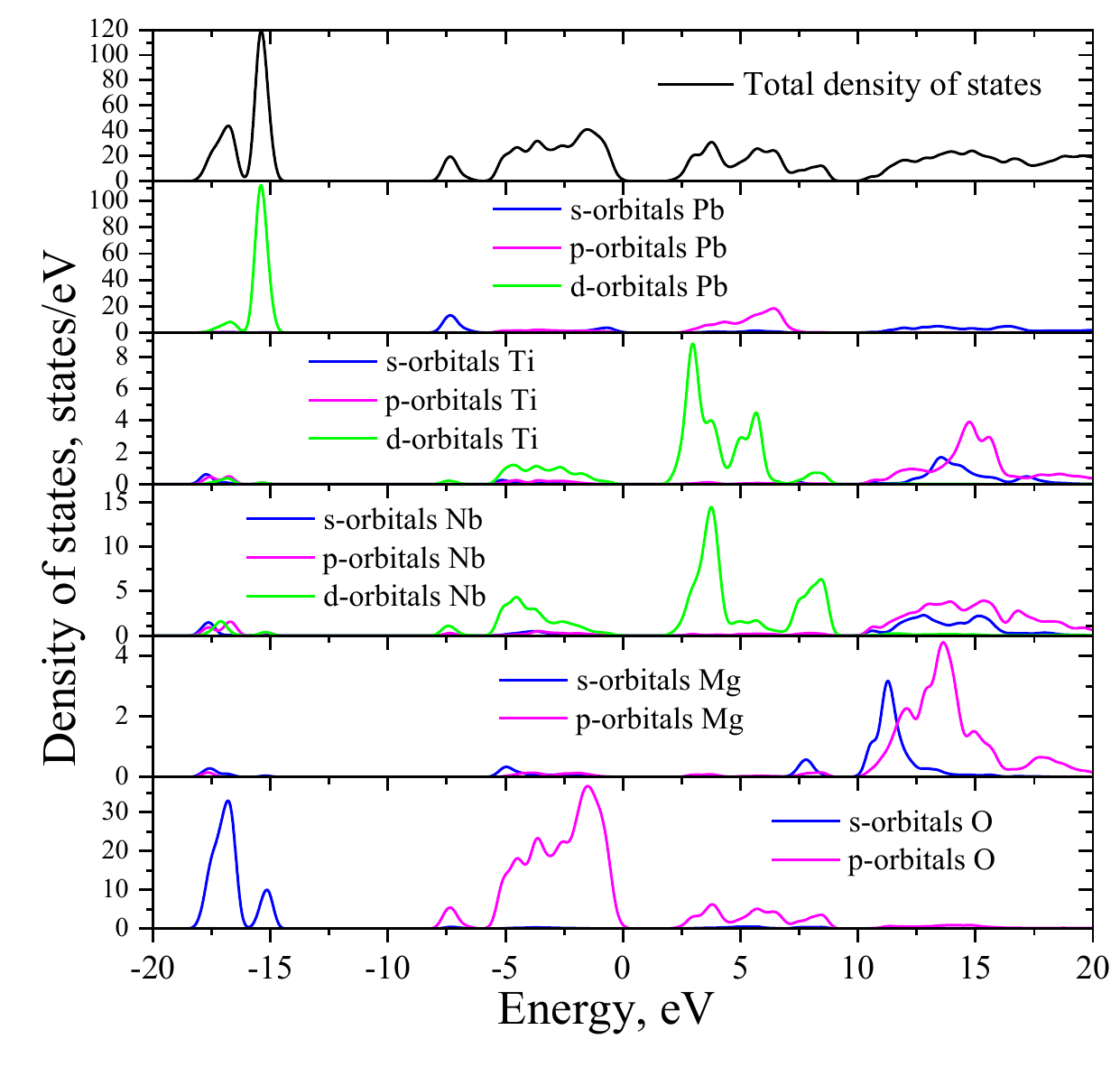}}
\caption{(Colour online) The DOS and PDOS for the 0.75PMN--0.25PT, calculated by the GGA(PBEsol) method.} \label{fig3}
\end{figure}

To establish the genetic origin of the electronic states of the
0.75PMN--0.25PT compound, the distributions of the total and partial
density of states (DOS and PDOS, respectively) were calculated and
presented in figure~\ref{fig3}.

PDOS shows that deep electronic states from $-$18 to $-$15 eV energy range are 
mainly derived from O 2\textit{s}-orbitals and Pb 5\textit{d}-orbitals. Similar behavior
was inherent to the PDOS of PbTiO\textsubscript{3} perovskites \cite{Derk2023}. The valence band from $-$9 to 0 eV is formed by the 2\textit{p} O states and the 6\textit{s} Pb states, while the hybridized 3\textit{d} Ti states, 4\textit{d} Nb states, and 6\textit{p} Pb states mainly contribute the conduction band in the energy range from 3.2 to 7.5 eV.
It should be noted that the most significant contribution near the top
of the valence band is observed from the 2\textit{p} orbitals of O ions. The
bottom of the conduction band is mainly formed by the 3\textit{d} and 4\textit{d} states
of Ti and Nb atoms, respectively (figure~\ref{fig3}). In general, the obtained
distribution of the density of states corresponds to previous
theoretical calculations \cite{Tan2018}, and we also see a good agreement
with the results obtained separately for components of the PMN--PT solid
solution \cite{Derk2023, Yang2006}.

\begin{table}[h!]
\caption{Influence of the Hubbard $U$ parameter values (eV) chosen for the $3d$ Ti, $4d$ Nb, and $2p$ O orbitals on the bandgap values (${E}_{g}$, eV) and volume per formula unit ($V$,~\AA$^{3}$/f.u.).}
\label{tab2}
\vspace{2ex}
\begin{center}
\renewcommand{\arraystretch}{0}
\begin{tabular}{|c|c|c|c|c||c|c|c|c|c|}
\hline
\multicolumn{3}{|c|}{Hubbard $U$ parameters} & ${E}_{g}$ & $V$ &\multicolumn{3}{|c|}{Hubbard $U$ parameters} & ${E}_{g}$ & $V$ \strut\\
\cline{1-3}\cline{6-8}
 $U_{d,\rm{Ti}}$ & $U_{d,\rm{Nb}}$ & $U_{p,\rm{O}}$ & & & $U_{d,\rm{Ti}}$ & $U_{d,\rm{Nb}}$ & $U_{p,\rm{O}}$ & &\strut\\
 \hline
0 & 0 & 0 & 2.13 & 65.88 & 9  & 8 & 0 & 2.53 &  65.45 \strut\\
\hline
5  & 3 & 0& 2.29 & 65.72 & 9  & 8 & 5 & 3.15 & 65.30  \strut\\
\hline
6  & 4 & 0& 2.50 & 65.64 & 9  & 8 & 6 & 3.19 & 65.23 \strut\\
\hline
7  & 5 & 0& 2.52 & 65.58 & 9  & 9 & 5 & 3.19 & 65.18 \strut\\
\hline
8  & 6 & 0& 2.56 &  65.52 & 9  & 9 & 6 & 3.24 & 65.16 \strut\\
\hline
8 & 6 & 2& 2.82 & 65.46 & 10  & 8 & 0 & 2.52 & 65.37  \strut\\
\hline
8  & 6 & 4 & 2.93 & 65.40 & 10  & 8 & 1 & 2.71 & 65.26 \strut\\
\hline
8  & 6 & 5& 2.97 & 65.31 & 10  & 8 & 2 & 2.89 & 65.21 \strut\\
\hline
8  & 6 & 6& 3.01 &  65.26 & 10  & 8 & 3 & 3.06 & 65.15  \strut\\
\hline
8  & 7 & 5& 3.00 &  65.23 & 10  & 8 & 4 & 3.19 & 65.11 \strut\\
\hline
8  & 8 & 5 & 3.04 & 65.18 & 10  & 8 & 5 & 3.24 & 65.07  \strut\\
\hline
8  & 8 & 6 & 3.08 &  65.16 & 10  & 8 & 6 & 3.29 & 65.05 \strut\\
\hline
 \end{tabular}
\renewcommand{\arraystretch}{1}
\end{center}
\end{table}

\begin{figure}[htb]
\centerline{\includegraphics[width=0.75\textwidth]{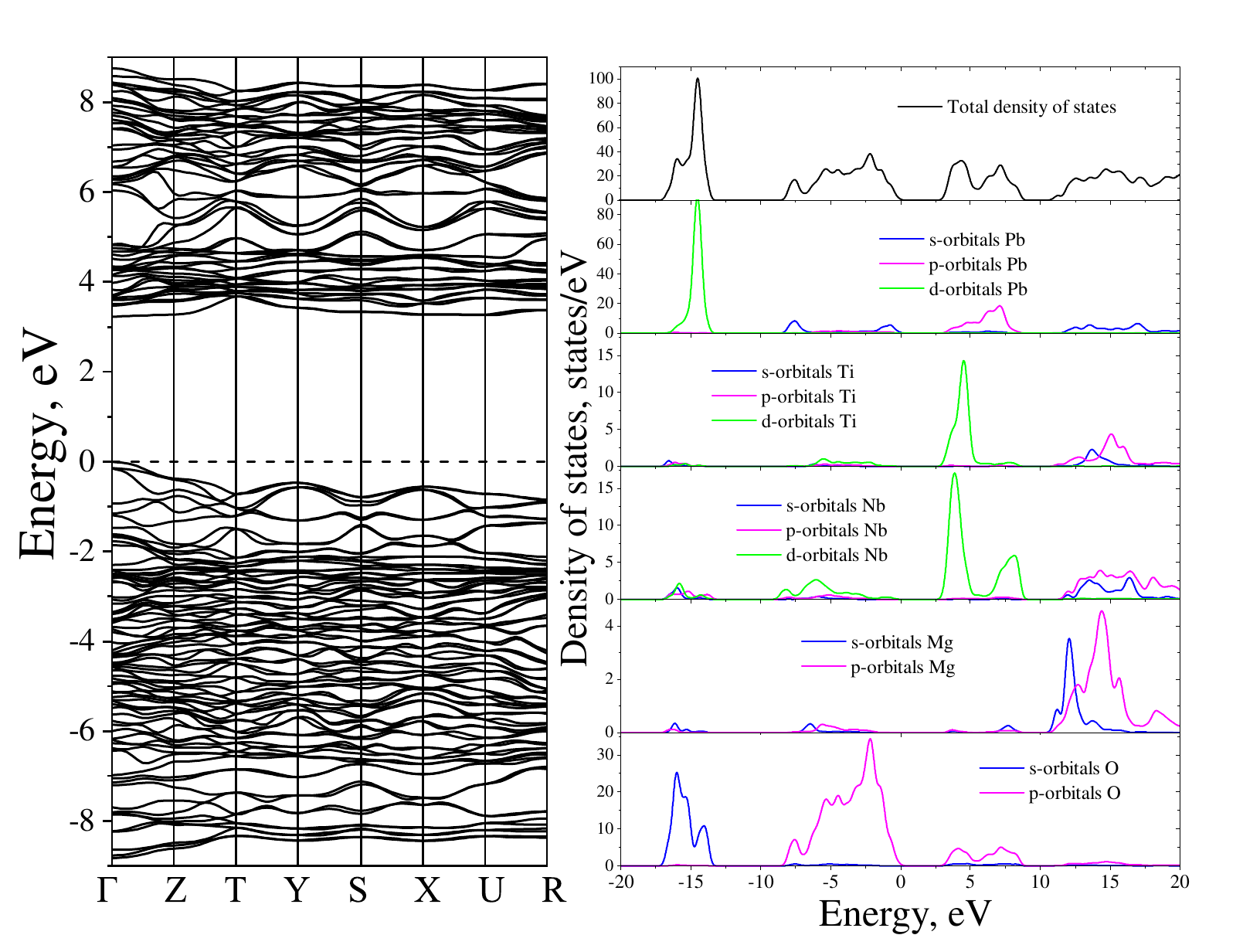}}
\caption{(Colour online) Band structure (left), DOS, and PDOS (right) of the 0.75PMN--0.25PT compound, calculated by the GGA+U method.} \label{fig4}
\end{figure}

The next stage of our investigation was based on analyzing the genesis
of the electronic states and their energy position and selecting the
Hubbard $U$ parameters to obtain more accurate electronic structures for
the 0.75PMN--0.25PT system. First, we chose $U$ parameters for the
\textit{d}-orbitals of Ti and Nb ions ($U_{d,\rm{Ti}}$,
$U_{d,\rm{Nb}}$), which are considerable for the
formation of electronic states near the bottom of the conduction band
and did not apply $U_{\rm{d}}$ for Pb \textit{d}-orbitals because it does not
influence the variation in bandgap energy due to its deep position in
the valence band of the electronic structure of PMN--PT system (see figure~\ref{fig3}). Calculations showed that taking into account $U$ corrections only for Ti
and Nb atoms refines the value of the bandgap by approximately 20\% (the largest ${E}_{g} = 2.56$~eV obtained for $U_{d,\rm{Ti}} = 8$~eV, $U_{d,\rm{Nb}} = 6$~eV, $U_{p,\rm{O}} = 0$~eV, see table~\ref{tab2}). To improve the obtained results, in addition to the $U_{d,\rm{Ti}}$, and $U_{d,\rm{Nb}}$ parameters, the non-zero $U$ parameter for oxygen atoms ($U_{p,\rm{O}}$) was taken into account. This approach was used for simple oxide semiconductors (TiO\textsubscript{2}, ZnO) \cite{Park2010, Kovalenko2021, Bovgyra2019, Bovg2019}, and for perovskite-type ABO\textsubscript{3} oxide crystals, particularly for PbTiO\textsubscript{3} \cite{Derk2023} and BaTiO\textsubscript{3} \cite{Derkaoui2023}. The obtained results for different sets of $U$ parameters are presented in table~\ref{tab2}. Consideration of the three Hubbard $U$ parameters allows us to achieve the bandgap value of 3.24 eV with the parameters $U_{d,\rm{Ti}} = 10$~eV, $U_{d,\rm{Nb}} = 8$~eV, and $U_{p,\rm{O}} = 5$~eV,
which is in excellent agreement with the experimental data \cite{Wan2004}, and the obtained corresponding electronic spectrum for 0.75PMN--0.25PT is presented in figure~\ref{fig4}. The results also showed that for the structural properties such as the value of the volume per formula unit, the GGA+U yielded the results (65.07~\AA$^3$/f.u.) close to the experimental values when $U_{d,\rm{Ti}} = 10$~eV, $U_{d,\rm{Nb}} = 8$~eV, and $U_{p,\rm{O}} = 5$~eV. It should be noted that the received set is not unique, we obtained another set that gives an experimental value of the bandgap: $U_{d,\rm{Ti}} = 9$~eV, $U_{d,\rm{Nb}} = 9$~eV, and $U_{p,\rm{O}} = 6$~eV (see table~\ref{tab2}), but the value of the volume per formula unit is 65.16~\AA$^{3}$/f.u. Therefore, all further calculations were carried out for $U$ parameters: $U_{d,\rm{Ti}} = 10$~eV, $U_{d,\rm{Nb}} = 8$~eV, and $U_{p,\rm{O}} = 5$~eV.

Following the implementation of $U$ parameters, the band structure
exhibits an increased dispersion compared to the outcome without their
inclusion. The PDOS analysis showed that the bands associated with
contributions from \textit{d} orbitals of Ti and Nb shift up the energy
scale, thereby widening the bandgap.

To determine the bonding nature of Pb--O, Ti--O, Mg--O, and Nb--O, the
bond lengths, bond population, and Mulliken charges for 0.75PMN--0.25PT
system were calculated using the GGA(PBEsol)+U method. The results are
presented in table~\ref{tab3}.

\begin{table}[htb]
\caption{Calculated average bond length (\textit{d},~\AA), bond population ($e$), and Mulliken charges [$q(e)$] of 0.75PMN--0.25PT system using GGA(PBEsol) and GGA(PBEsol)+U methods.}
\label{tab3}
\vspace{2ex}
\begin{center}
\renewcommand{\arraystretch}{0}
\begin{tabular}{|c|c|c|c|c|c|c|}
\hline
 Bond & \textit{d} &\multicolumn{2}{c|}{bond population}& Atom &\multicolumn{2}{c|}{$q(e)$}\strut\\
\cline{3-4} \cline{6-7}
& & GGA & GGA+U& & GGA & GGA+U \strut\\
\hline
 Pb--O & 2.70 &$ - $0.08 &  $ - $0.11 & Pb & 1.22 & 1.41 \strut\\
 \hline
Mg--O & 2.11 & $ - $0.65 & $ - $0.64 & Mg & 1.61 & 1.63\strut\\
\hline
Ti--O & 2.00 & 0.51 & 0.52 & Ti & 1.03 & 1.19 \strut\\
\hline
Nb--O & 2.01 &0.55 & 0.59 & Nb & 1.10  & 1.23 \strut\\
\hline
O--O & 2.82 &$ - $0.03& $ - $0.02 & O & $ - $0.80 & $ - $0.91\strut\\
\hline
\end{tabular}
\renewcommand{\arraystretch}{1}
\end{center}
\end{table}

The bond lengths of Pb--O, Mg--O, Ti--O, Nb--O and O--O are 2.70, 2.11,
2.00, 2.01, and 2.82~\AA, respectively, for the 0.75PMN--0.25PT compound.
It is necessary to note that the obtained results for bond lengths of
Pb--O, Ti--O, and O--O correspond to appropriate bond lengths received
theoretically and experimentally in PbTiO\textsubscript{3} perovskite
\cite{Derk2023, Shirane1956}.

The Mulliken charge analysis in a crystal lattice describes the degree
of charge transfer between ions and helps to establish the bonding type between ions.
Table~\ref{tab3} shows the calculated values of charge transfer $q(e)$ for all ions in the 0.75PMN--0.25PT compound. Based on the obtained results, we can conclude
that Ti and O atoms form covalent bonding as well as between Nb and O
atoms. However, there is no observation of covalent bonding between Mg
and O atoms, while the Pb--O bond shows a weak
covalent bonding. Our calculations found that the B-site atoms,
particularly Ti, Nb, and Mg, in relaxor ferroelectric 075PMN--PT have
different bonding characteristics with O atoms: Ti--O and Nb--O bonds are
strongly covalent, while Mg--O bond remains highly ionic, representing the 
absence of covalent bonding with O atoms. The received results indicate that
the different bond characteristics might contribute to the relaxor properties
of the PMN--PT solid solution. Such bond behavior confirms the 
theoretical results obtained earlier for PMN relaxor \cite{Yang2006}. 

Next, we investigated the optical properties of PMN--PT compounds because
these materials are considered to be promising for photovoltaic applications.
In particular, we calculated the real ($\varepsilon_{1}$) and
imaginary ($\varepsilon_{2}$) parts of the dielectric function
and, based on them, calculated the absorption coefficient~($\alpha$),
which is directly related to them. The optical properties of the
material can be described using the complex dielectric function $\varepsilon $($\omega$),
which has two components --- real $\varepsilon_{1}$($\omega$) and
imaginary~$\varepsilon_{2}(\omega)$~\cite{Ambr2006}:
\begin{align}
\label{delta-def}
\varepsilon(\omega) = \varepsilon_{1}(\omega) + \ri\varepsilon_{2}(\omega).
\end{align}
Usually, the electronic structure is directly related to the imaginary
part of the dielectric function [$\varepsilon_{2}(\omega)$] and
indicates all possible transitions from filled to unfilled states. The
value of $\varepsilon_{2}(\omega)$ was calculated from the
expression:
\begin{align}
\label{delta-def}
\varepsilon_{2}(\omega) = \ \frac{2e^{2}\piup}{\Omega\varepsilon_{0}}\sum_{k,v,c}^{}{\left|\left\langle \psi_{k}^{c}\left| \widehat{u\ } \times r \right|\psi_{k}^{v} \right\rangle\right|\delta(E_{k}^{c} - E_{k}^{v} - E)},
\end{align}
where $ \Omega $ is the unit cell volume, \textit{u} is the polarization
vector, \(\psi_{k}^{c}\) and \(\psi_{k}^{v}\) are the wave functions of
the valence and conduction bands, respectively. The real part of the
dielectric function $\varepsilon_{1}$($\omega$) can be calculated using
the Kramers--Kronig relation, starting from $\varepsilon_{2}(\omega)$
\cite{Ambr2006}:
\begin{align}
\label{delta-def}
\varepsilon_{1}(\omega) = 1 + \frac{2}{\piup}\mathcal{P}\!\!\int_{0}^{\infty}{\frac{\omega^{\prime}\varepsilon_{2}(\omega^{\prime})}{\omega^{\prime \, 2} - \omega^{2}}\rd\omega^{\prime}},
\end{align}
where $\mathcal{P}$ denotes the fundamental value of the integral.

\begin{figure}[htb]
\centerline{\includegraphics[width=0.85\textwidth]{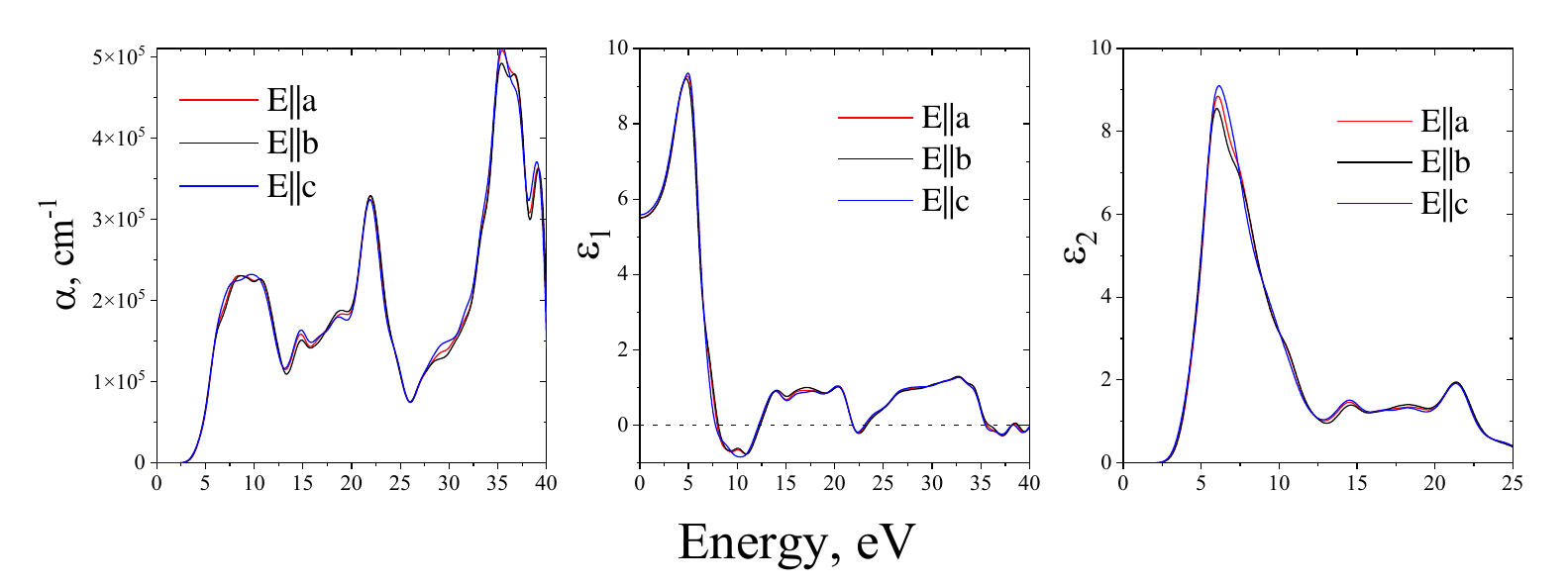}}
\caption{(Colour online) Optical spectra for different light polarizations of 0.75PNM--0.25PT calculated by the GGA(PBEsol)+U method.} \label{fig5}
\end{figure}

The absorption coefficient $\alpha$($\omega$) is directly related to the
dielectric function and can be calculated using the following
expression \cite{Adachi2009}:
\begin{align}
\label{delta-def}
\alpha(\omega) = \sqrt{2\omega}\left[\sqrt{\varepsilon_{1}^{2}(\omega) + \varepsilon_{2}^{2}(\omega)} - \varepsilon_{1}(\omega)\right]^{1/2}.
\end{align}
The optical properties of the 0.75PMN--0.25PT system are presented in
figure~\ref{fig5}, calculated using the GGA(PBEsol)+U approximation. The figure
shows a small anisotropy of the optical properties observed in the
spectra.

The spectra of the imaginary part of the dielectric function $\varepsilon_{2}$ are directly related to the electronic band structure and describe the light
absorption of the compounds. The prominent peaks of the $\varepsilon_{2}$ spectra are indicated approximately by 6.14, 14.53, 18.16, and 21.33 eV (see figure~\ref{fig5}, right plot). The transition from 2\textit{p} O [valence band maximum (VBM)] to 3\textit{d} Ti and 4\textit{d} Nb [conduction band minimum~(CBM)] orbitals are mainly
represented by the peak near 6.14 eV. By contrast, the transition from
2\textit{p} O (VBM) to 6\textit{p} Pb (CB) orbitals is indicated by a 14.53 eV peak. In
addition, the optical peaks near 18.16 and 21.33~eV are associated with
internal electronic excitation transitions of the 5\textit{d} Pb and 2\textit{s} O states
near the valence band to the semi-core states of the conduction band. It
should be noted that every peak in the $\varepsilon_{2}$ function is not always related to a single interband transition because the electronic band structure
may include many direct or indirect transitions with the same energy
corresponding to the identical peak. The real part of the dielectric
function ($\varepsilon_{1}$) received similar results to the imaginary part, where the calculated zero frequency limit [$\varepsilon_{1}(0)$] for 0.75PMN--0.25PT compound is 7.94, 8.06, and 7.73 for different light polarizations (figure~\ref{fig5}, middle plot) in the case of GGA(PBEsol)+U method.

The initial value in the absorption spectrum $\alpha$($\omega$) lies near 3.06 eV for
0.75PMN--0.25PT solid solution and reaches its maximum value of 35.22 eV (figure~\ref{fig5}, left plot).
Thus, essential differences in optical parameters in the energy range
from 3 to 35 eV make 0.75PMN--0.25PT solid solution very useful for applications in optical devices.

\section{Conclusions}

Within the GGA(PBEsol)+U method, the structural, electronic, and optical
properties of the 0.75PNM--PT compound were studied. The equilibrium
lattice parameters and displacement from the high-symmetry perovskite
position of Pb ions and B-cations were established. Using the Hubbard
corrections method makes it possible to overcome the usual error in GGA
calculations and obtain the band gap value corresponding to the
experiment. Different bonding behavior between ions is established in
0.75PNM--PT compound: Ti--O and Nb--O bonds are strongly covalent;
Pb--O bonds are weakly covalent, and Mg--O bonds are exclusively ionic.
Based on the optical spectra of the real and imaginary parts of the
dielectric function of 0.75PNM--PT solid solution, an analysis of
interband transitions has been carried out. Accordingly, these results
can be used by scientists to expand the targeted area of the PMN--PT
compound.

\section*{Acknowledgements}

This work supported by the Ministry of Education and Science of Ukraine.



\begin{thebibliography}{10}
\bibitem{Luo2000}  Luo~H., Xu~G., Xu~H., Wang~P., Yin~Z.,  Jpn. J. Appl. Phys., 2000, \textbf{39}, 5581, \doi{10.1143/JJAP.39.5581}.
\bibitem{Kutnjak2006} Kutnjak~Z., Petzelt~J., Blinc~R., Nature, 2006, \textbf{441}, 956, \doi{10.1038/nature04854}.  
\bibitem{Alguero2006}  Alguer\'o~M., Moure~A., Pardo~L., Holc~J., Kosec~M., Acta Mater., 2006, \textbf{54}, 501--511,\\ \doi{10.1016/j.actamat.2005.09.020}.
\bibitem{Bokov2000} Bokov~A.~A., Ye~Z.~G., Appl. Phys. Lett., 2000,  \textbf{77}, 1888, \doi{10.1063/1.1310629}.
\bibitem{Noheda2002} Noheda~B., Cox~D.~E., Shirane~G., Gao~J., Ye~Z.~G., Phys. Rev. B,  2002, \textbf{66}, 054104,\\ \doi{10.1103/PhysRevB.66.054104}.
\bibitem{Ye2002} Ye~Z.~G., Curr. Opin. Solid State Mater. Sci., 2002, \textbf{6}, 35--44,  \doi{10.1016/S1359-0286(02)00019-0}.
\bibitem{Semak2023} Semak~S., Kapustianyk~V., Eliyashevskyy~Yu., Bovgyra~O., Kovalenko~M.,  Mostovoi~U., Doudin~B., Kundys~B., J.~Phys.:~Condens.~Matter, 2023, \textbf{35}, 094001, \doi{10.1088/1361-648X/aca579}.
\bibitem{Tu2006} Tu~C.~S., Wang~F.~T., Chien~R.~R., Schmidt~H.~V., Hung~C.~M., Tseng~C.~T., Appl.~Phys.~Lett., 2006, \textbf{88}, 032902,\\ \doi{10.1063/1.2165278}.
\bibitem{Liew2022} Liew~W.~H., Chen~Y., Alexe~M., Yao~K., Small, 2022, \textbf{18}, 2106275, \doi{10.1002/smll.202106275}.
\bibitem{Makh2019} Makhort~A.~S., Schmerber~G., Kundys~B.,  Mater. Res. Express, 2019, \textbf{6}, 066313,  \doi{10.1088/2053-1591/ab0758}.
\bibitem{Makh2018} Makhort~A.~S., Chevrier~F., Kundys~D., Doudin~B., Kundys~B., Phys. Rev. Mater., 2018, \textbf{2}, 012401,\\ \doi{10.1103/PhysRevMaterials.2.012401}.
\bibitem{Bokov2002} Bokov~A.~A., Ye~Z.~G., Phys. Rev. B, 2002, \textbf{66}, 094112, \doi{10.1103/PhysRevB.66.094112}. 
\bibitem{Shvar2013}  Shvartsman~V.~V., Kholkin~A.~L., Raevski~I.~P., Raevskaya S.~I., Savenko~F.~I., Emelyanov~ A.~S., J. Appl. Phys., 2013, \textbf{113}, 187208, \doi{10.1063/1.4801964}. 
\bibitem{Li2019} Li~J., Yin~R., Su~X., Wu~H.~H., Li~J., Qin~S., Sun~S., Chen~J., Su~Y., Qiao~L., Guo~D., Bai~Y., Acta Mater., 2020, \textbf{182}, 250--256, \doi{10.1016/j.actamat.2019.11.017}.  
\bibitem{Li2021} Li~J., Li J., Wu~H.~H., Zhou~O., Chen~J., Lookman~T., Su~Y., Qiao L.,  Bai Y., ACS Appl. Mater. Interfaces, 2021, \textbf{13}, 38467--38476,  \doi{10.1021/acsami.1c07714}.
\bibitem{Tan2018} Tan~T., Takenaka~H., Xu~C., Duan~W., Grinberg~I., Rappe~A.~M., Phys. Rev. B, 2018, \textbf{97}, 174101,\\ \doi{10.1103/PhysRevB.97.174101}. 
\bibitem{Li2020} Li~C., Xu~B., Lin~D., Zhang~S., Bellaiche~L., Shrout~T.~R., Li~F., Phys. Rev. B, 2020, \textbf{101}, 140102(R),\\ \doi{10.1103/PhysRevB.101.140102}.
\bibitem{Grin2004} Grinberg~I., Rappe~A.~M., Phys. Rev. B, 2004, \textbf{70}, 220101(R), \doi{10.1103/PhysRevB.70.220101}.  
\bibitem{Takenaka2014}  Takenaka~H., Grinberg~I., Shin~Y.~H., Rappe~A.~M., Ferroelectrics, 2014, \textbf{469}, 1--13,\\ \doi{10.1080/00150193.2014.948341}.
\bibitem{Grin2005} Grinberg~I., Suchomel~M.~R., Davies~P.~K., Rappe~A.~M., J. Appl. Phys., 2005, \textbf{98}, 094111, \doi{10.1063/1.2128049}
\bibitem{Grinb2004} Grinberg~I., Cooper~V.~R., Rappe~A.~M., Phys. Rev. B, 2004,  \textbf{69}, 144118, \doi{10.1103/PhysRevB.69.144118}.
\bibitem{Clark2005}	Clark~S.~J., Segall~M.~D., Pickard~C.~J., Hasnip~P.~J., Probert~M.~I.~J., Refson~K., Payne~M.~C., Z. Kristallogr., 2005, \textbf{220}, 567--570, \doi{10.1524/zkri.220.5.567.65075}.
\bibitem{Bovgyra2015} Bovgyra~O.~V., Kovalenko~M.~V., In: Proceedings of the Conference ``2015 International Young Scientists Forum on Applied Physics'' (Dnipropetrovsk, 2015), IEEE, New York, 2015, 1--4, \doi{10.1109/YSF.2015.7333157}.
\bibitem{Bovgyra2023} Bovgyra~O., Kozachenko~O., Kovalenko~M., Kapustianyk~V., Appl. Nanosci., 2023, \textbf{13}, 5003--5010,\\ \doi{10.1007/s13204-022-02662-9}. 
\bibitem{Bovgyra2016} Bovgyra~O.~V., Kovalenko~M.~V., J. Nano- Electron. Phys., 2016, \textbf{8}, 02031, \doi{10.21272/jnep.8(2).02031}.
\bibitem{Kapus2022} Kapustianyk~V., Semak~S., Chornii~Yu., Bovgyra~O., Kovalenko~M., Physica B, 2022, \textbf{639}, 413929,\\ \doi{10.1016/j.physb.2022.413929}.
\bibitem{Davies2000} Davies~P.~K., Akbas~M.~A., J. Phys. Chem. Solids, 2000, \textbf{61}, 159--166, \doi{10.1016/S0022-3697(99)00275-9}.
\bibitem{Sepl2011} Sepliarsky~M., Cohen~R.~E., J. Phys.: Condens. Matter, 2011, \textbf{23}, 435902, \doi{10.1088/0953-8984/23/43/435902}.
\bibitem{Makh2022} Makhort~A., Gumeniuk~R., Dayen~J.~F., Dunne~P., Burkhardt~U.,
Viret~M., Doudin~B., Kundys~B., Adv.~Opt.~Mater., 2022, \textbf{10}, 2102353, \doi{10.1002/adom.202102353}.
\bibitem{Grinb2007} Grinberg~I., Rappe~A.~M., Phase Transitions, 2007, \textbf{80}, 351--368, \doi{10.1080/01411590701228505}.
\bibitem{Zhang2017} Zhang~Y., Sun~J., Perdew~J.~P., Wu~X., Phys. Rev. B, 2017, \textbf{96}, 035143, \doi{10.1103/PhysRevB.96.035143}. 
\bibitem{Wan2004} Wan~X., Chan~H.~L.~W., Choy~C.~L., Zhao~X., Luo~H., J. Appl. Phys., 2004, \textbf{96}, 1387, \doi{10.1063/1.1767287}.
\bibitem{Derk2023} Derkaoui~I., Achehboune~M., Eglitis~R.~I., Popov~A.~I., Rezzouk~A., Materials, 2023, \textbf{16}, 4302,\\  \doi{10.3390/ma16124302}.
\bibitem{Yang2006} Yang~K., Wang~C.~L., Li~J.~C., Integr. Ferroelectr., 2006, \textbf{78}, 113--117, \doi{10.1080/10584580600660033}.
\bibitem{Park2010} Park~S.~G., Magyari-K\"ope~B., Nishi~Y., Phys. Rev. B, 2010, \textbf{82}, 115109, \doi{10.1103/PhysRevB.82.115109}.
\bibitem{Kovalenko2021} Kovalenko~M., Bovgyra~O., Franiv~A., Dzikovskyi~V., Mater. Today: Proc., 2021, \textbf{35}, 604--608,\\ \doi{10.1016/j.matpr.2019.11.274}. 
\bibitem{Bovgyra2019} Bovgyra~O., Kovalenko~M., Dzikovskyi~V., Moroz~M., In: Proceedings of the Conference ``2019 IEEE 2nd Ukraine Conference on Electrical and Computer Engineering (UKRCON)'' (Lviv, 2019), IEEE, 2019, 726--731, \doi{10.1109/UKRCON.2019.8879928}.
\bibitem{Bovg2019} Bovgyra~O., Kovalenko~M., Bovhyra~R., Dzikovskyi~V., J. Phys. Stud., 2019, \textbf{23}, 4301, \doi{10.30970/jps.23.4301}.
\bibitem{Derkaoui2023}
Derkaoui~I., Achehboune~M., Boukhoubza~I., El~Adnani~Z., Rezzouk~A., Comput. Mater. Sci., 2023, \textbf{217}, 111913, \doi{10.1016/j.commatsci.2022.111913}.
\bibitem{Shirane1956} Shirane~G., Pepinsky~R., Frazer~B.~C., Acta Crystallogr., 1956, \textbf{9}, 131--140, \doi{10.1107/S0365110X56000309}.
\bibitem{Ambr2006} Ambrosch-Draxl~C., Sofo~J.~O., Comput. Phys. Commun., 2006, \textbf{175},  1--14, \doi{10.1016/j.cpc.2006.03.005}.
\bibitem{Adachi2009} Adachi~S., Properties of Semiconductor Alloys: Group-IV, III-V and II-VI Semiconductors, John Wiley \& Sons, 2009.


\end{thebibliography}

%
%
\newpage

\ukrainianpart

\title{Локальна структура, електронні та оптичні властивості Pb(Mg$_{1/3}$Nb$_{2/3}$)O$_3$-PbTiO$_3$: першопринципне дослідження}
%
%
  \author{М. Коваленко, О. Бовгира, В. Капустяник, О. Козаченко}
  \address{Кафедра фізики твердого тіла, Фізичний факультет, Львівський національний університет імені Івана Франка,  вул. Кирила і Мефодія 8, 79005 Львів, Україна }

\makeukrtitle

\begin{abstract}
\tolerance=3000%
Кристали на основі перовскіту Pb(Mg$_{1/3}$Nb$_{2/3}$)O$_3$-PbTiO$_3$ викликають значний науковий інтерес завдяки своїм цікавим властивостям і перспективі використання в п’єзоелектричних та фотовольтаїчних пристроях. Щоб зрозуміти вплив локальної структури на  властивості таких матеріалів, у цій роботі ми зосередилися на вивченні структурних, електронних і оптичних властивостей Pb[(Mg$_{1/3}$Nb$_{2/3}$)$_{0.75}$Ti$_{0.25}$]O$_3$ в межах теорії функціоналу густини. Результати розрахунків показали, що використання апроксимації GGA(PBEsol) для оптимізації структури дає добре узгодження з експериментальними даними. Завдяки підбору параметрів Габбарда $U$ для функціоналу GGA(PBEsol) ми досягли ширини забороненої зони для Pb[(Mg$_{1/3}$Nb$_{2/3}$)$_{0.75}$Ti$_{0.25}$]O$_3$, яка добре узгоджується з експериментальною. Дослідження заселеностей зв’язків показало, що зв’язок Mg--O не ковалентний, тоді як існує значний ковалентний зв’язок між  {Ti--O} та Nb--O. Такий різний характер зв’язування між атомами має відповідати за релаксорні властивості сполуки Pb[(Mg$_{1/3}$Nb$_{2/3}$)$_{0.75}$Ti$_{0.25}$]O$_3$. Крім того, було проведено дослідження оптичних властивостей Pb[(Mg$_{1/3}$Nb$_{2/3}$)$_{0.75}$Ti$_{0.25}$]O$_3$ шляхом застосування поправок Габбарда  $U$ для усунення похибки наближення GGA  та підтверджено аналіз електронного спектру.
\keywords локальна структура, заборонена зона, густина станів, оптичні властивості

\end{abstract}

\end{document}